\documentclass[a4paper,11pt]{article}

\usepackage{comment}
\usepackage{pos}
\usepackage{lipsum} 
\usepackage{etoolbox}
\usepackage{subcaption}
\usepackage{enumitem}
\usepackage{lineno}
\patchcmd{\thebibliography}
  {\labelsep}
  {\itemsep 0pt \labelsep}
  {}{}

\title{IceCat-2: Updated IceCube Event Catalog of Alert Tracks}

\ShortTitle{IceCat-2: Updated IceCube Event Catalog of Alert Tracks}

\author{The IceCube Collaboration \\{\normalsize \normalfont(a complete list of authors can be found at the end of the proceedings)}\\}

\emailAdd{azegarelli@icecube.wisc.edu}
\emailAdd{anna.franckowiak@icecube.wisc.edu}
\emailAdd{sommani.giacomo@icecube.wisc.edu}
\emailAdd{nora.valtonen-mattila@icecube.wisc.edu}
\emailAdd{tianlu.yuan@icecube.wisc.edu}

\abstract{
We present preliminary results for IceCat-2, the second public catalog of IceCube Alert Tracks, which plans to build and improve upon the first release, IceCat-1. The initial catalog, last updated in October 2023, included all real-time alerts issued since 2016, as well as events observed by IceCube since the start of full-detector data collection in 2011 that would have triggered an alert if the program had been in place at that time. IceCat-2 plans to expand on this by incorporating all additional alerts since IceCat-1, and reprocessing all events with significantly improved reconstruction algorithms. A key advancement in IceCat-2 will come from an updated reconstruction technique introduced by the IceCube Collaboration in September 2024. This approach substantially enhances the angular resolution of muon track alerts, while also improving statistical coverage. With respect to IceCat-1, the 50\%(90\%) angular uncertainty on track alerts is expected to be reduced by a factor of approximately 5(4). These refined reconstructions will allow us to revisit possible correlations between past alerts and sources in gamma-ray and X-ray catalogs. The enhanced precision may uncover new astrophysical associations with known astrophysical sources, offering deeper insight into potential cosmic ray accelerators.

\vspace{4mm}

{\bfseries Corresponding authors:}
Angela Zegarelli$^{1*}$,
Anna Franckowiak$^{1}$, 
Giacomo Sommani$^{1}$, 
Nora Valtonen-Mattila$^{1}$, 
Tianlu Yuan$^{2}$\\
{$^{1}$ \itshape Ruhr-Universit{\"a}t Bochum, Germany}\\
{$^{2}$ \itshape University of Wisconsin–Madison, USA}\\[4mm]
$^*$ Presenter
}

\ConferenceLogo{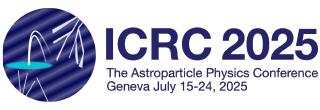}

\FullConference{39th International Cosmic Ray Conference (ICRC2025)\\
 15–24 July 2025\\
Geneva, Switzerland\\}

\begin{document}

\maketitle
\vspace{-4mm}
\section{Introduction}
\vspace{-3mm}
High-energy neutrinos play a crucial role in multimessenger astronomy by providing unique insights into cosmic ray (CR) accelerators. Unlike gamma ($\gamma$) rays, which can originate from both leptonic and hadronic processes, neutrinos are unambiguous signatures of hadronic interactions. Produced through hadronuclear and photohadronic processes, they travel unimpeded by matter or magnetic fields and preserve directional information about their sources, making them ideal messengers for identifying distant astrophysical accelerators. Because of their extremely small interaction cross section, their detection is highly challenging, requiring kilometer-scale Cherenkov detectors embedded in natural transparent media such as ice or water. These detectors capture Cherenkov light emitted by secondary particles produced when neutrinos interact with nuclei and travel faster than the local speed of light in the medium.

The IceCube Neutrino Observatory at the South Pole~\cite{icecube}, currently the largest operating neutrino telescope, made the first detection of a diffuse astrophysical neutrino flux in 2013~\cite{icecube_diffuse}. Since then, IceCube has provided strong evidence for extreme CR accelerators, likely of extragalactic origin, although no dominant source class has been conclusively identified. In the multimessenger framework, combining neutrino detections with electromagnetic (EM) and gravitational-wave observations is essential for locating and characterizing the sources of high-energy particles. Neutrinos offer precise timing and directional information that can guide follow-up searches for EM counterparts. Moreover, Cherenkov-based neutrino detectors continuously monitor the sky and can promptly notify the scientific community of interesting events.

IceCube has been operating a real-time alert system since 2016~\cite{Aartsen:2016nxy}. Although the system includes multiple event types, the focus here is on track-like alerts. These events originate from charged-current interactions between muon neutrinos and quarks in the ice nuclei. Due to their excellent angular resolution, which improves with energy, track-like alerts are particularly valuable for neutrino astronomy. 
The system allows to rapidly notify the multimessenger community about likely astrophysical neutrino track-like events detected by IceCube, initiating follow-ups to detect source candidates. A major breakthrough occurred in September 2017, when a neutrino with an energy of approximately 300~TeV (IC-170922A) was found in spatial and temporal coincidence with the flaring blazar TXS 0506+056. This association, supported by $\gamma$-ray data from \textit{Fermi}-LAT, had a significance of $\sim$3$\sigma$\cite{txs}. Archival analysis revealed a possible earlier neutrino flare from the same source between September 2014 and March 2015, with a significance of $3.5\sigma$ independent of the 2017 alert~\cite{txs_prior}. These observations demonstrate the potential of real-time multi-messenger astronomy.

In 2019, the real-time framework for track-like events was upgraded to improve event selection and alert message content~\cite{blaufuss_2019}. Following this upgrade, IceCube released its first catalog of alert events, named \textit{IceCat-1} \cite{icecat1}. This catalog includes all real-time alerts issued since 2016, and earlier events since 2011 that would have triggered an alert under the current system, and it was periodically updated until October 2023. A recent enhancement was introduced in September 2024 \cite{2024GCN.37625....1I}. IceCube is adopting a revised follow-up reconstruction strategy that applies different algorithms based on the event’s energy deposition or topology. This hybrid approach enhances angular resolution across a broad energy range and ensures robust statistical coverage of the reported localization uncertainties. Thus, an updated catalog, named \textit{IceCat-2}, is now under processing. 

Sec.~\ref{sec:icecat2} presents a preliminary overview of the IceCat-2 catalog, highlighting refined event selection criteria, the inclusion of newly processed alerts, and improvements in reconstruction. In relation to this, Sec.~\ref{sec:comparison_icecat1_icecat2} provides a comparison with IceCat-1, focusing on directional estimates and associated uncertainties. In Sec.~\ref{sec:coincidences_with_sources}, we assess whether previously identified coincidences with candidate neutrino sources remain valid under the new reconstruction. Preliminary results from a cross-correlation with known gamma-ray catalogs are presented in Sec.~\ref{sec:catalog_cross_check}. A summary is given in Sec.~\ref{sec:summary}.
\vspace{-4mm}
\section{IceCat-2: Updated IceCube Event Catalog of Alert Tracks}
\label{sec:icecat2}
\vspace{-3mm}
We present a preliminary overview of the IceCat-2 catalog, which updates and extends the previous IceCat-1 collection of likely astrophysical neutrino track-like events detected by IceCube. 

The main features of IceCat-2 are:
\begin{itemize}[noitemsep, topsep=0.1pt]
\item Reconstructed directions for all IceCat-1 events have been updated using the latest reconstruction algorithms, implemented in the real-time alert system since September 2024.
\item Events likely caused by atmospheric muons or other background sources have been excluded from the sample. Since October 2022, a \textit{veto} method has been used to reject atmospheric muons that might pass the alert selection criteria. This veto utilizes IceTop, an array of ice-Cherenkov tanks located on the surface, to detect CR-induced air showers accompanying track-like events observed in the ice. As a result, the refined IceCat-1 dataset contains 340 events out of the original 348\footnote{IceCat-1 included vetoed events because the cosmic-ray veto criteria were introduced after the catalog compilation.}.
\item Inclusion of additional track-like alerts since the last IceCat-1 update in October 2023, extending the catalog coverage up to January 2, 2025 (IC-250102A). This results in 25 additional events compared to the last IceCat-1 update. Alerts occurring before September 29, 2024 were originally reconstructed with older algorithms and have now been reprocessed using the latest reconstruction framework.
\item Reprocessing of all events (both the original IceCat-1 and the additional alerts) using the most up-to-date data processing framework and calibrations to ensure a uniform and refined dataset.
\end{itemize}
\begin{figure}[t!]
\centering
\includegraphics[width=0.85\linewidth]{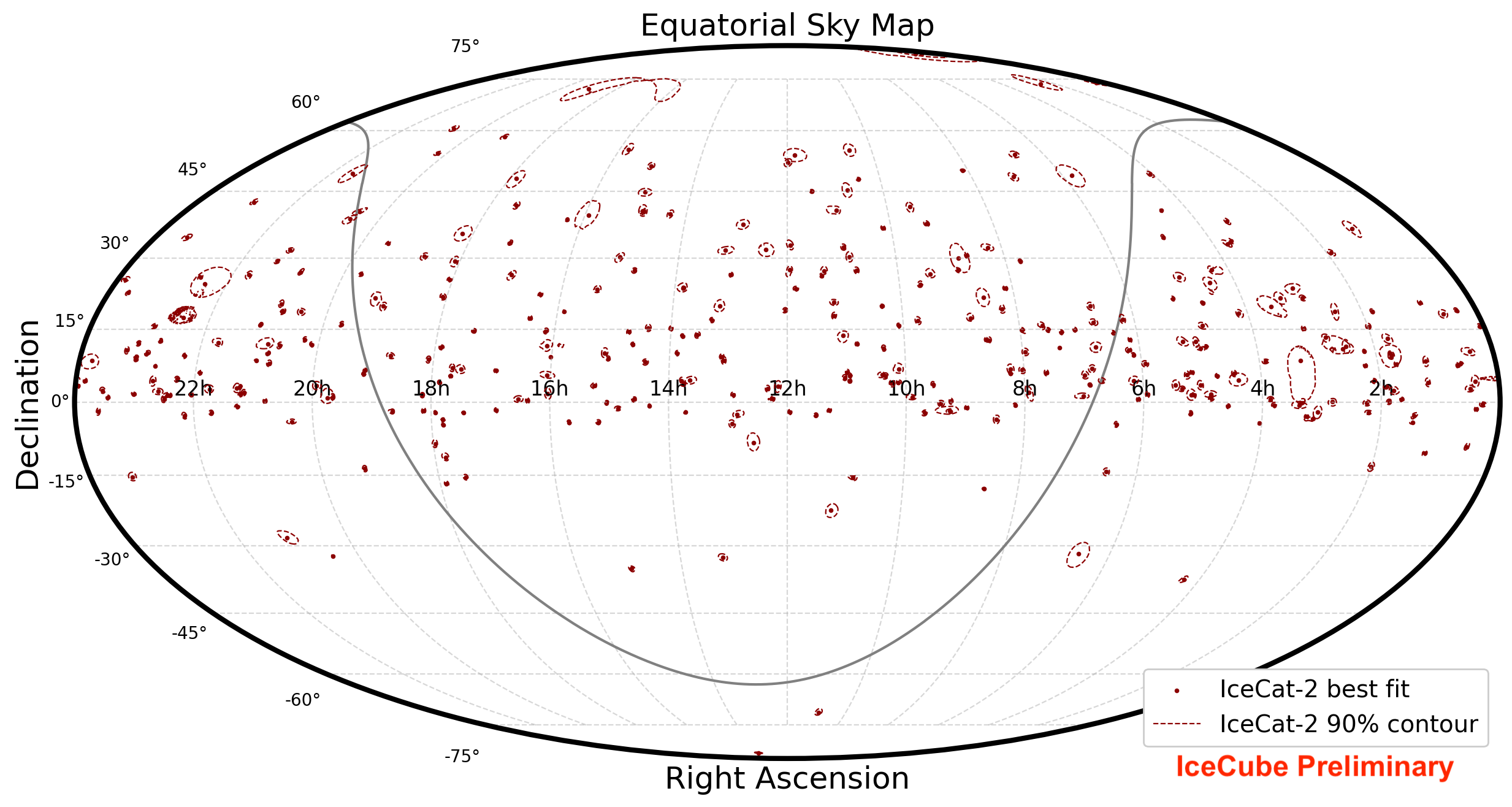}
\caption{The all-sky distribution of alerts from the IceCat-2 catalog in equatorial coordinates, including the corresponding 90\% containment contours. The thin line indicates the location of the Galactic Plane.}\label{fig:skymap}
\end{figure}
The preliminary IceCat-2 catalog includes 365 track-like alerts, with their sky distribution shown in Fig.~\ref{fig:skymap}. Issued at an average rate of 26.8 events per year, $\sim$9.9 and 17 fall into the Gold and Bronze channels, corresponding to average astrophysical probabilities of 50\% and 30\%, assuming a power-law spectrum with index 2.19. This selection, adopted in IceCat-1 and currently used in real-time alerts, is based on an earlier IceCube measurement. Future versions of the catalog will incorporate an updated spectral index value, reflecting the softer spectrum observed in muon track events by IceCube. The impact of this change will be investigated in forthcoming studies.

\vspace{-3mm}
\subsection{Comparison of IceCat-2 with IceCat-1}
\label{sec:comparison_icecat1_icecat2}
\vspace{-1mm}
Due to the updates introduced in the IceCat-2 processing pipeline, described in Sec.~\ref{sec:icecat2}, the new catalog significantly improved the directional reconstruction of neutrino events. For a detailed description of the updated reconstruction algorithms used in this work, we refer to \cite{splinempe2023,IceCube:2023asu,pos2025_reco}, which focuses on the new reconstruction approach and outlines improvements in directional precision and the treatment of systematic uncertainties. 
\begin{figure}[t!]
\centering
\includegraphics[width=0.6\linewidth]{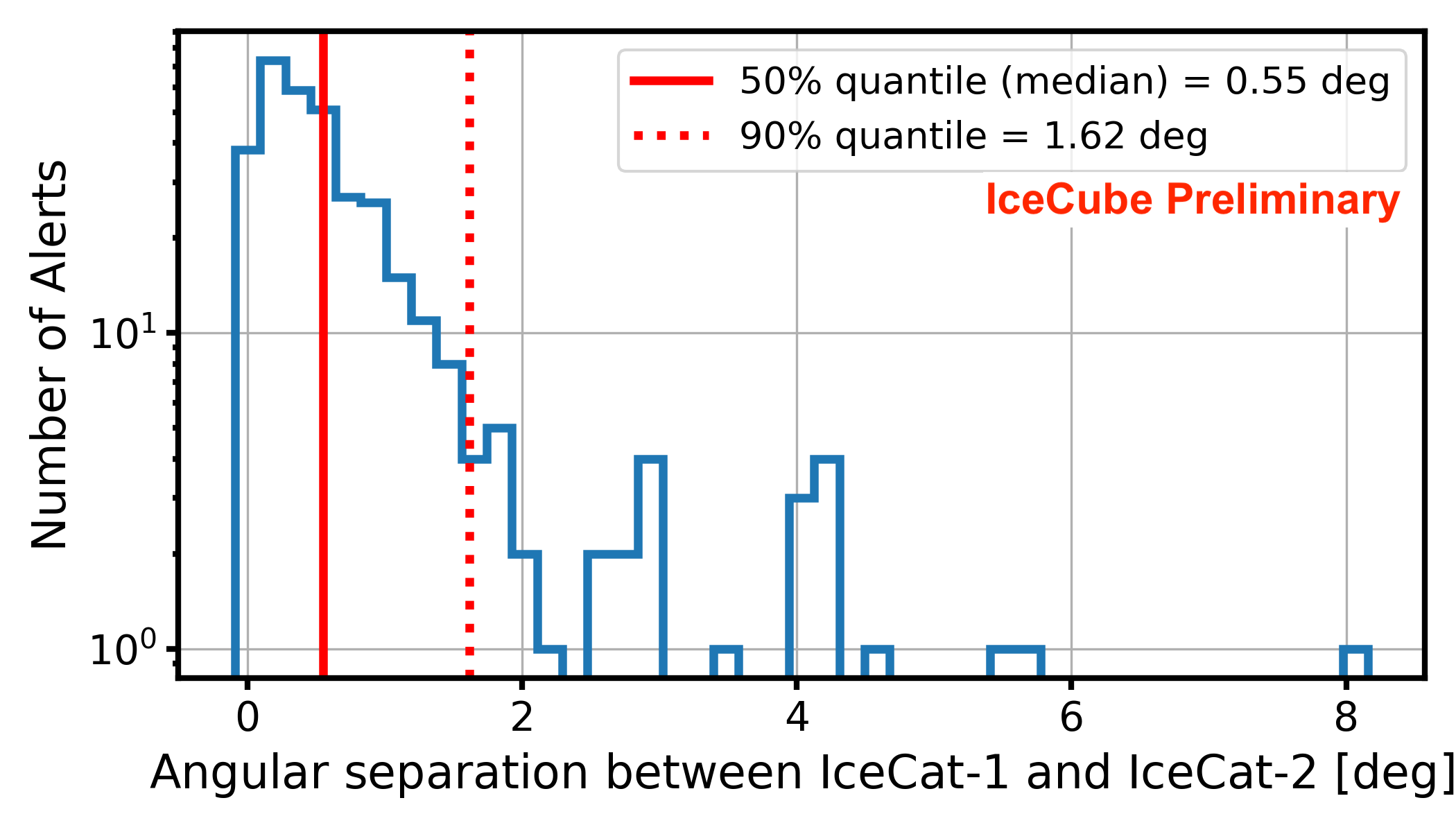}
\caption{Difference in reconstructed directions for IceCat-1 events reprocessed in the IceCat-2 sample (N = 340, excluding vetoed events). The red solid and dashed lines indicate the 50\% (median) and 90\% quantiles of angular separations, respectively.}\label{fig:angular_distances}
\end{figure}

\begin{figure}[h!]
\centering
\begin{subfigure}[t]{0.49\linewidth}
    \centering
    \includegraphics[width=\linewidth]{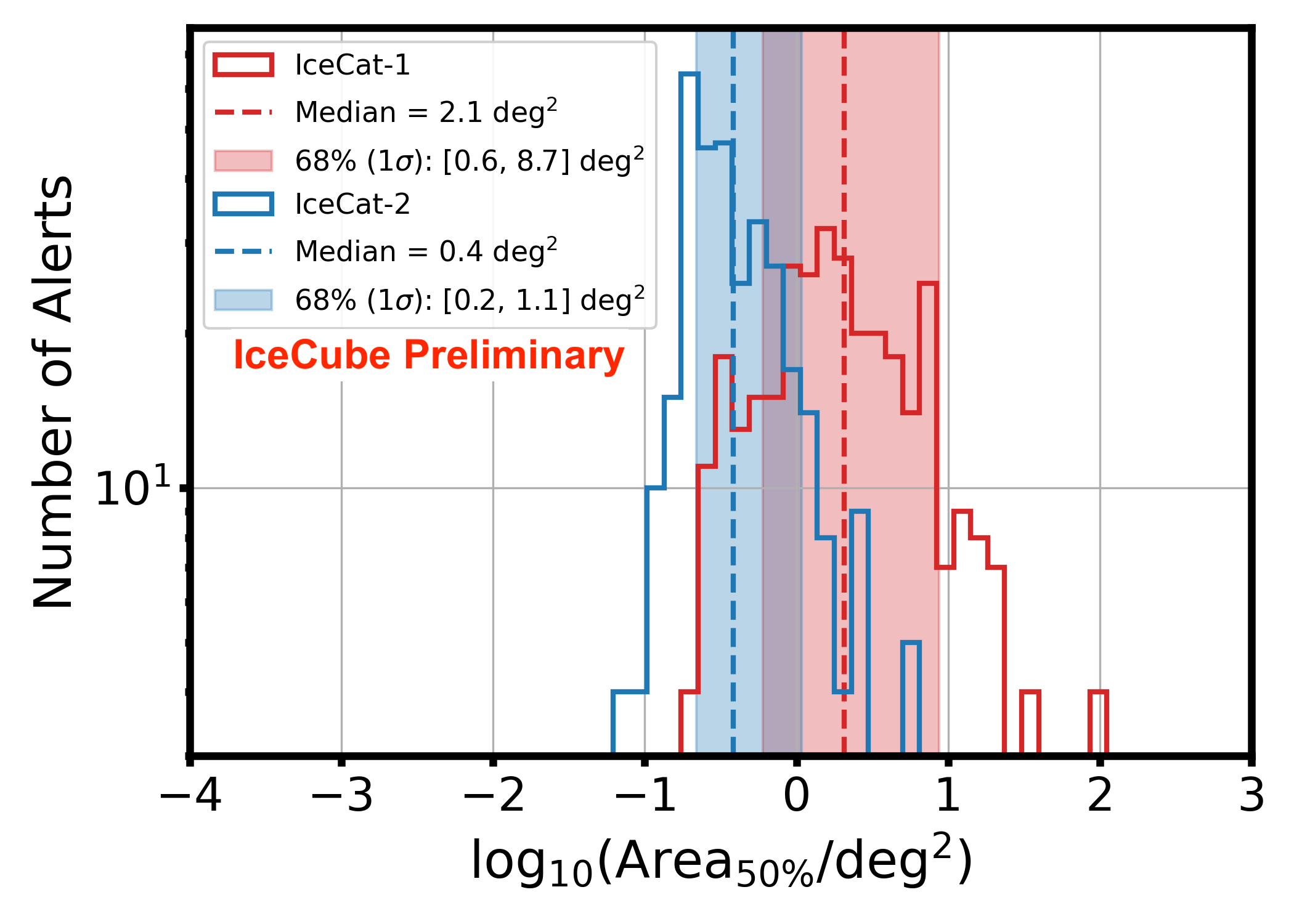}
    \caption{} \label{fig:area50_comparison}
\end{subfigure}
\hfill
\begin{subfigure}[t]{0.49\linewidth}
    \centering
    \includegraphics[width=\linewidth]{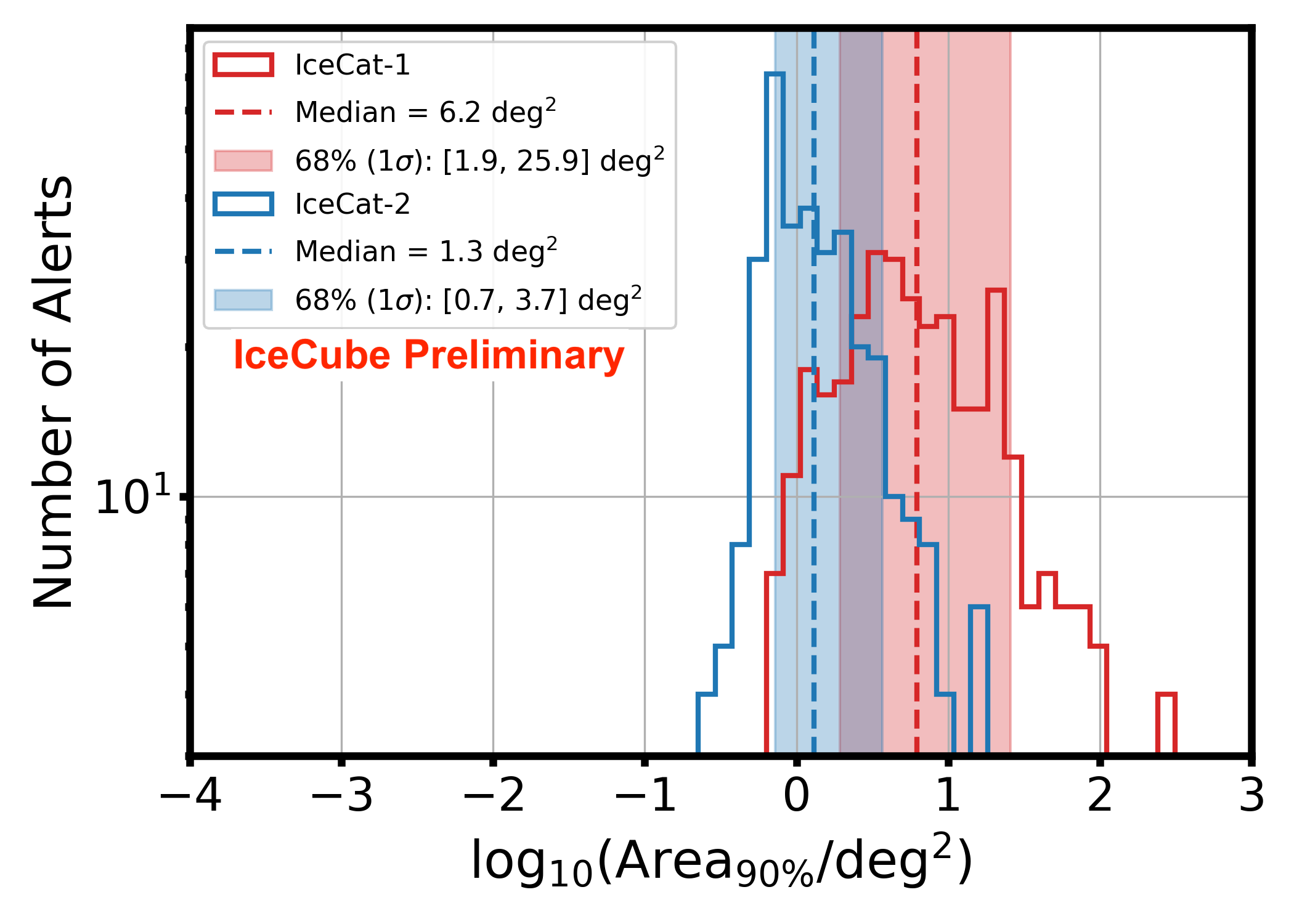}
    \caption{} \label{fig:area90_comparison}
\end{subfigure}
\begin{subfigure}[t]{0.5\linewidth}
    \centering
    \includegraphics[width=\linewidth]{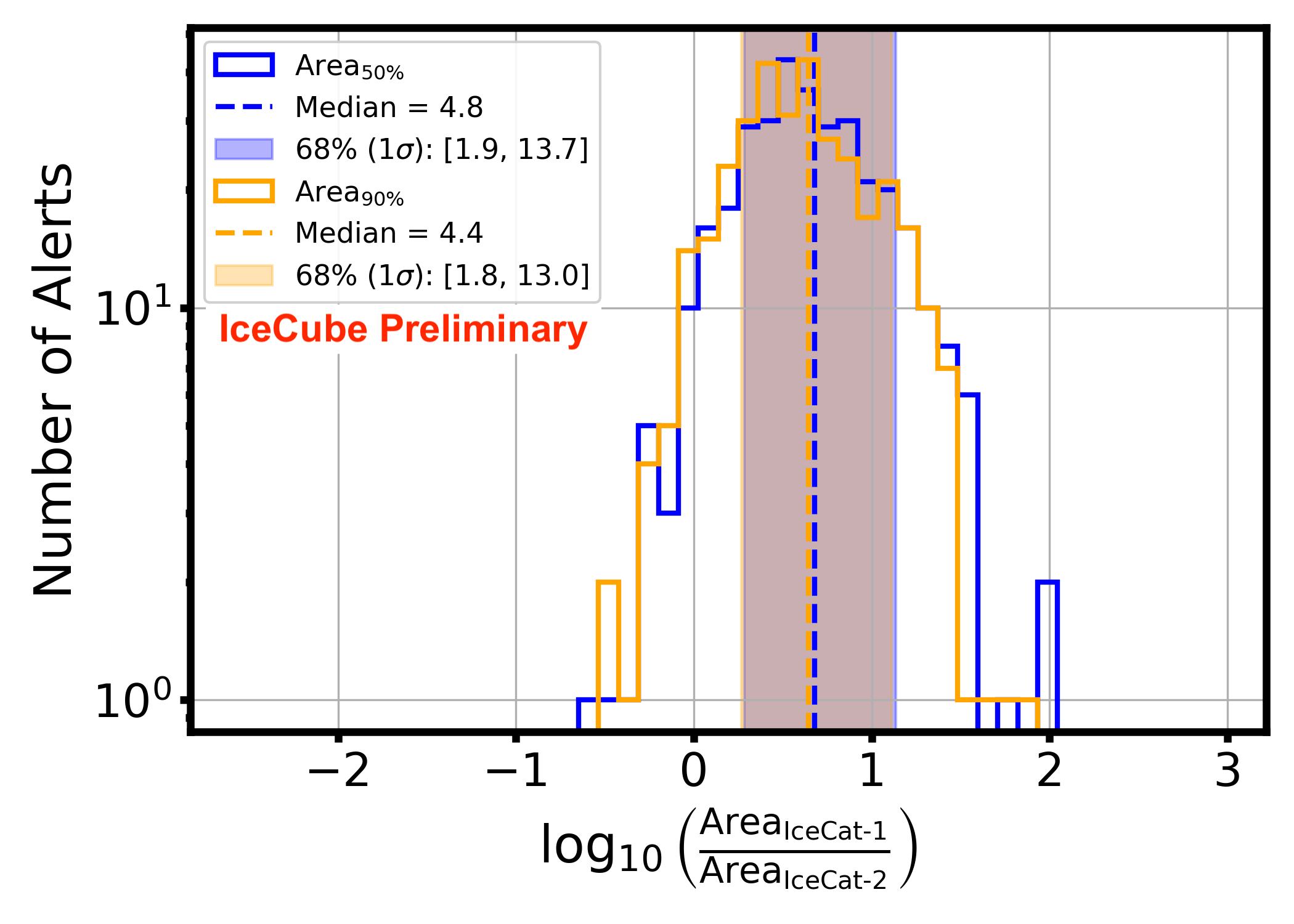}
    \caption{} \label{fig:ratio_comparison}
\end{subfigure}
\caption{Comparison of IceCat-1 and IceCat-2 containment performance:
(a)–(b): Distributions of 50\% and 90\% containment areas for IceCat-1 (red) and IceCat-2 (blue).
(c): Ratio of IceCat-1 to IceCat-2 areas at 50\% (blue) and 90\% (orange). Dashed lines mark medians, and shaded regions represent the corresponding 68\% quantile intervals.} 
\vspace{-4mm}
\label{fig:combined_comparison}
\end{figure}

Fig.~\ref{fig:angular_distances} shows the difference in reconstructed directions for IceCat-1 events reprocessed in the IceCat-2 sample, with 50(90)\% of the alerts with angular separation below 0.55(1.62) deg. The most notable improvement is the significant reduction in localization uncertainties around the best-fit direction, as illustrated in Fig.~\ref{fig:combined_comparison}. In particular, Figs.~\ref{fig:area50_comparison} and \ref{fig:area90_comparison} show that the median areas of the 50\% and 90\% containment contours in IceCat-2 are reduced by factors of approximately 5 and 4, respectively, compared to IceCat-1 (see also Fig.~\ref{fig:ratio_comparison}). Furthermore, the distributions are considerably narrower, with the spread around the median, quantified by the standard deviation $\sigma$, reduced by a factor between 7 and 9.
Small containment errors around best-fit directions are also evident in Fig.~\ref{fig:skymap}, which shows the all-sky distribution of alerts from the IceCat-2 catalog in equatorial coordinates, including the corresponding 90\% containment contours.

This substantial gain in angular precision directly enhances the efficiency of EM and multimessenger follow-up observations, increasing the chances of identifying astrophysical counterparts and advancing the search for CR accelerators.  In Sec.~\ref{sec:coincidences_with_sources}, we revisit the spatial coincidences with several potential high-energy neutrino sources previously identified through IceCube track alerts, now evaluated again in light of the updated reconstructions and improved uncertainty estimates.
\vspace{-4mm}
\section{Revision of spatial coincidences with relevant sources}
\label{sec:coincidences_with_sources}
\begin{figure}[t!]
\centering
\includegraphics[width=0.45\linewidth]{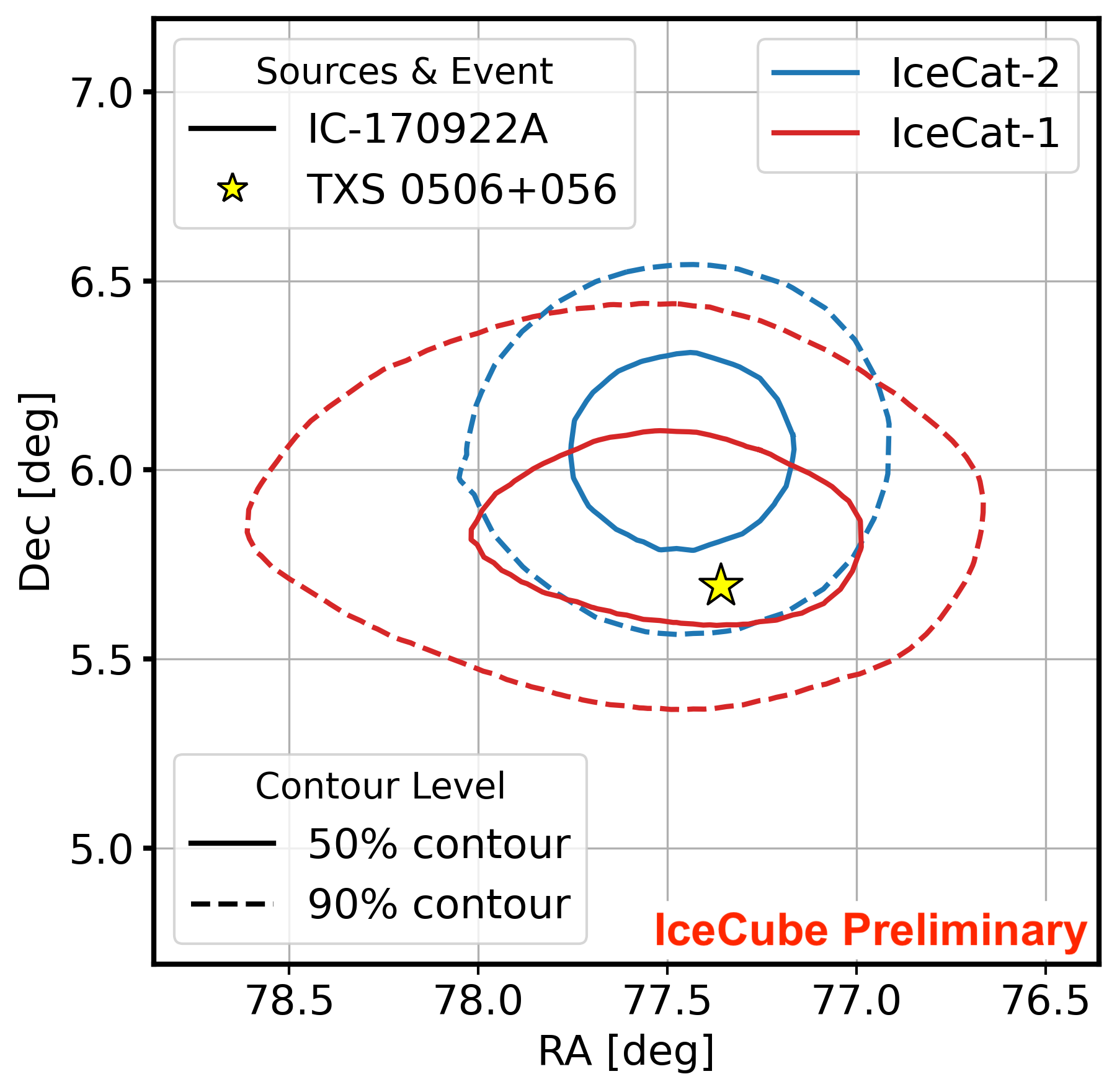}
\caption{Sky map showing spatial coincidences between IC-170922A and TXS0506+056 (yellow star). The tighter angular reconstruction contours highlight improvements in event localization from IceCat-1 (in red) to IceCat-2 (in blue), with solid and dashed lines representing the 50\% and 90\% containment regions, respectively.}\label{fig:txs}
\end{figure}

\vspace{-3mm}
Over the years, several candidate sources of high-energy astrophysical neutrinos have been identified through IceCube track-like alerts. Among the most prominent is TXS 0506+056, a blazar that gained significant attention following the detection of a high-energy neutrino (IC-170922A) in spatial and temporal coincidence with a gamma-ray flare from this source \cite{txs}. This marked the first compelling multimessenger association between a neutrino and an astrophysical object, providing evidence that blazars can act as CR accelerators. As shown in Fig.~\ref{fig:txs}, TXS 0506+056 remains spatially coincident with the revised IC-170922A localization, although it now lies within the 90\% containment region rather than the 50\% contour.
\begin{figure}[t!]
\centering
\begin{subfigure}[b]{0.49\linewidth}
    \includegraphics[width=\linewidth]{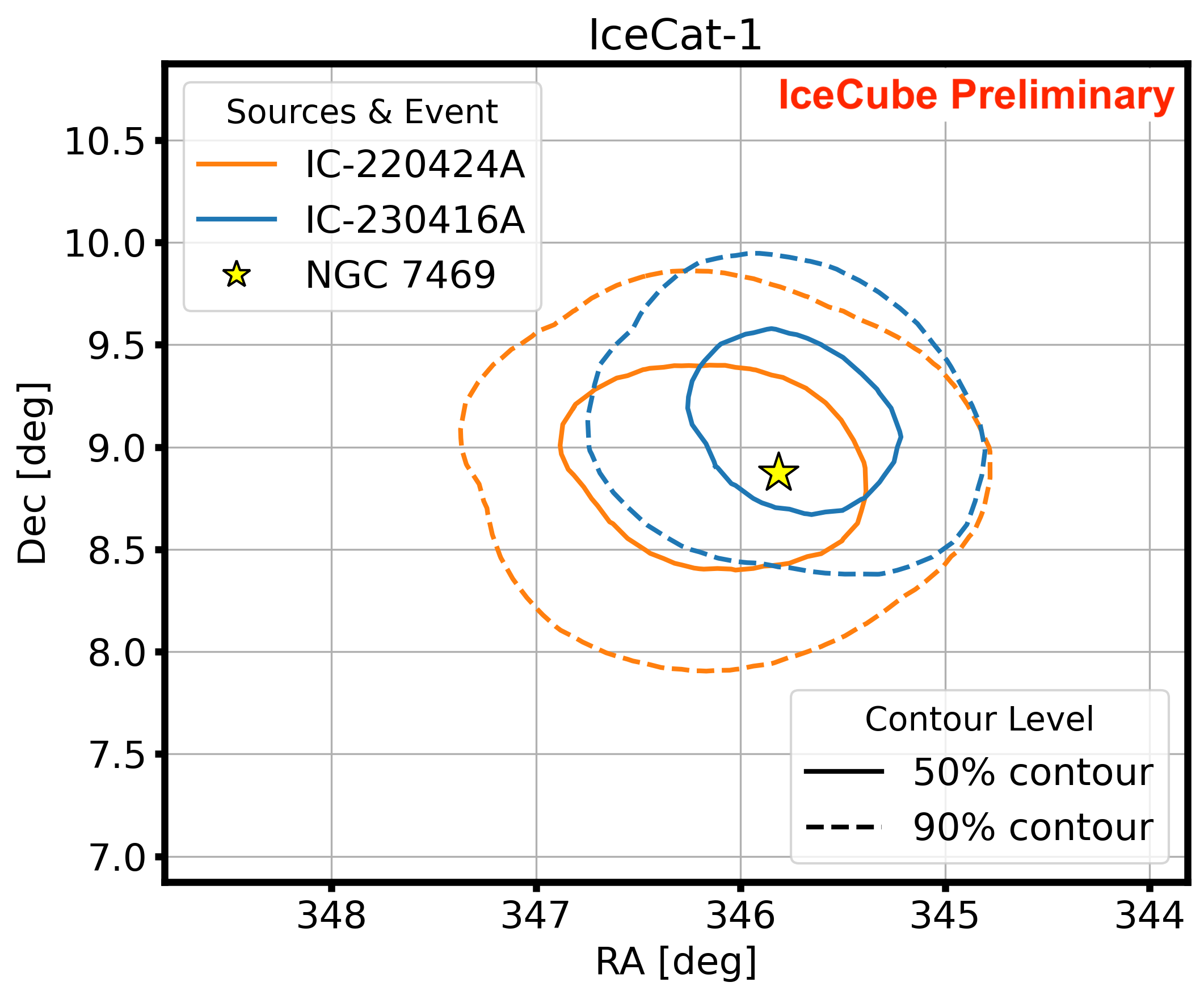}
    \caption{}
\end{subfigure}
\hfill
\begin{subfigure}[b]{0.49\linewidth}
    \includegraphics[width=\linewidth]{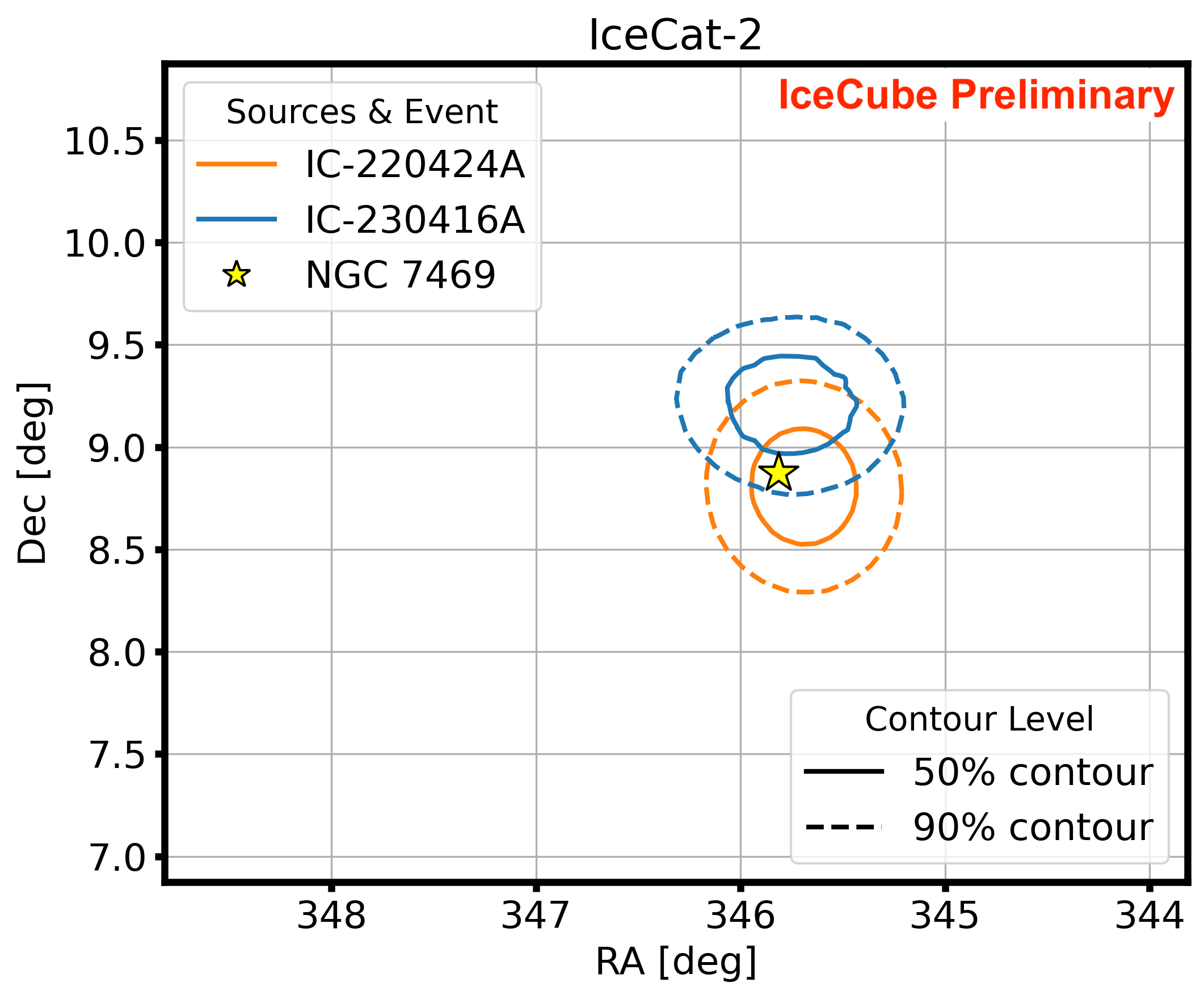}
    \caption{}
\end{subfigure}
\caption{Sky maps showing spatial coincidences between IC-220424A (in orange) and IC-230416A (in blue), and NGC 7469 (yellow star). Solid and dashed lines represent the 50\% and 90\% containment regions, respectively. Panels (a) and (b) display the contours of the two alerts in IceCat-1 and IceCat-2, respectively.}
\label{fig:ngc}
\end{figure}

\begin{figure}[h!]
\centering
\begin{subfigure}[t]{0.425\linewidth}
    \includegraphics[width=\linewidth]{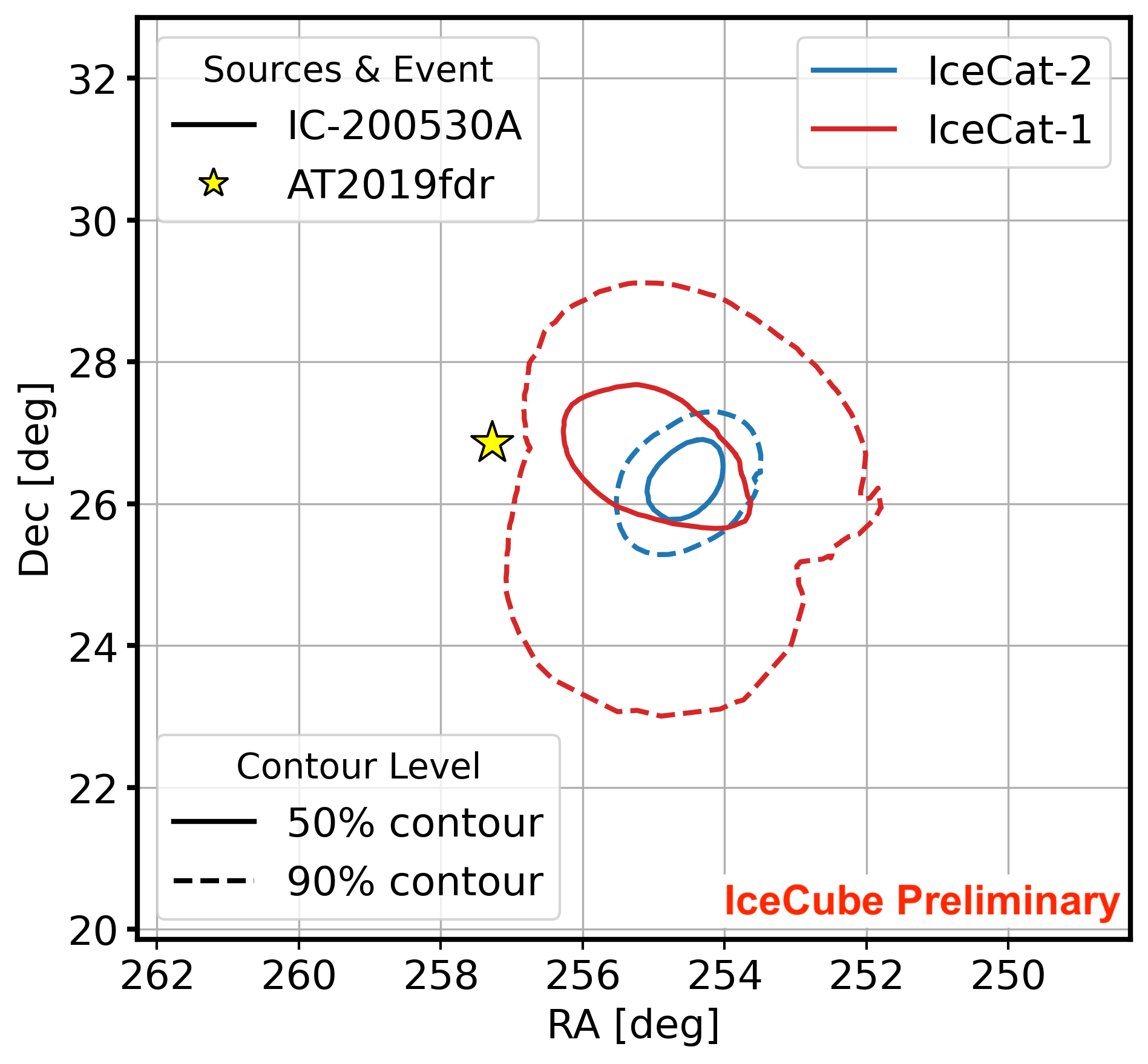}
    \caption{AT2019fdr}
\end{subfigure}
\hspace{0.01\linewidth}
\begin{subfigure}[t]{0.545\linewidth}
    \includegraphics[width=\linewidth]{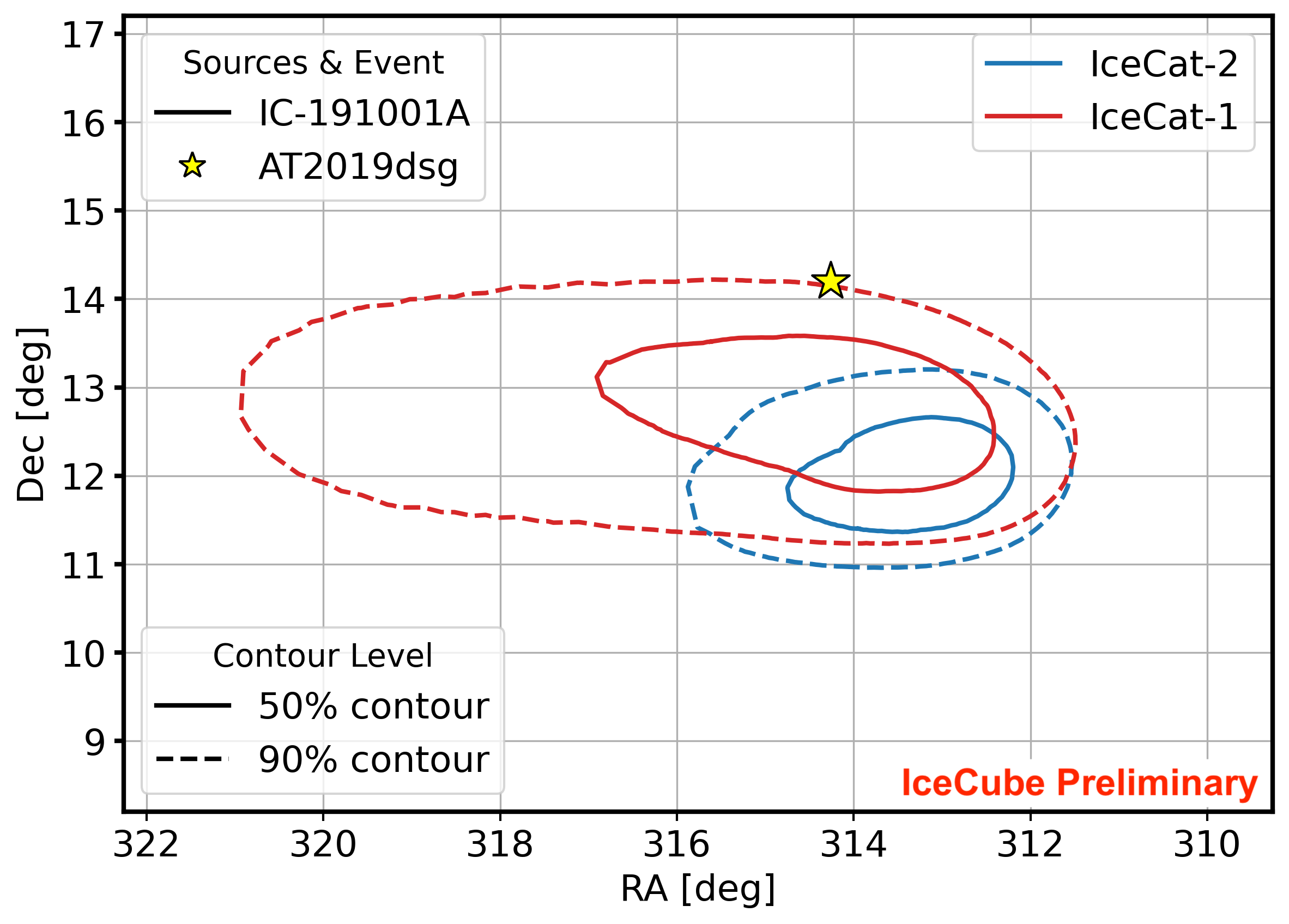}
    \caption{AT2019dsg}
\end{subfigure}

\begin{subfigure}[t]{0.43\linewidth}
    \centering
    \includegraphics[width=\linewidth]{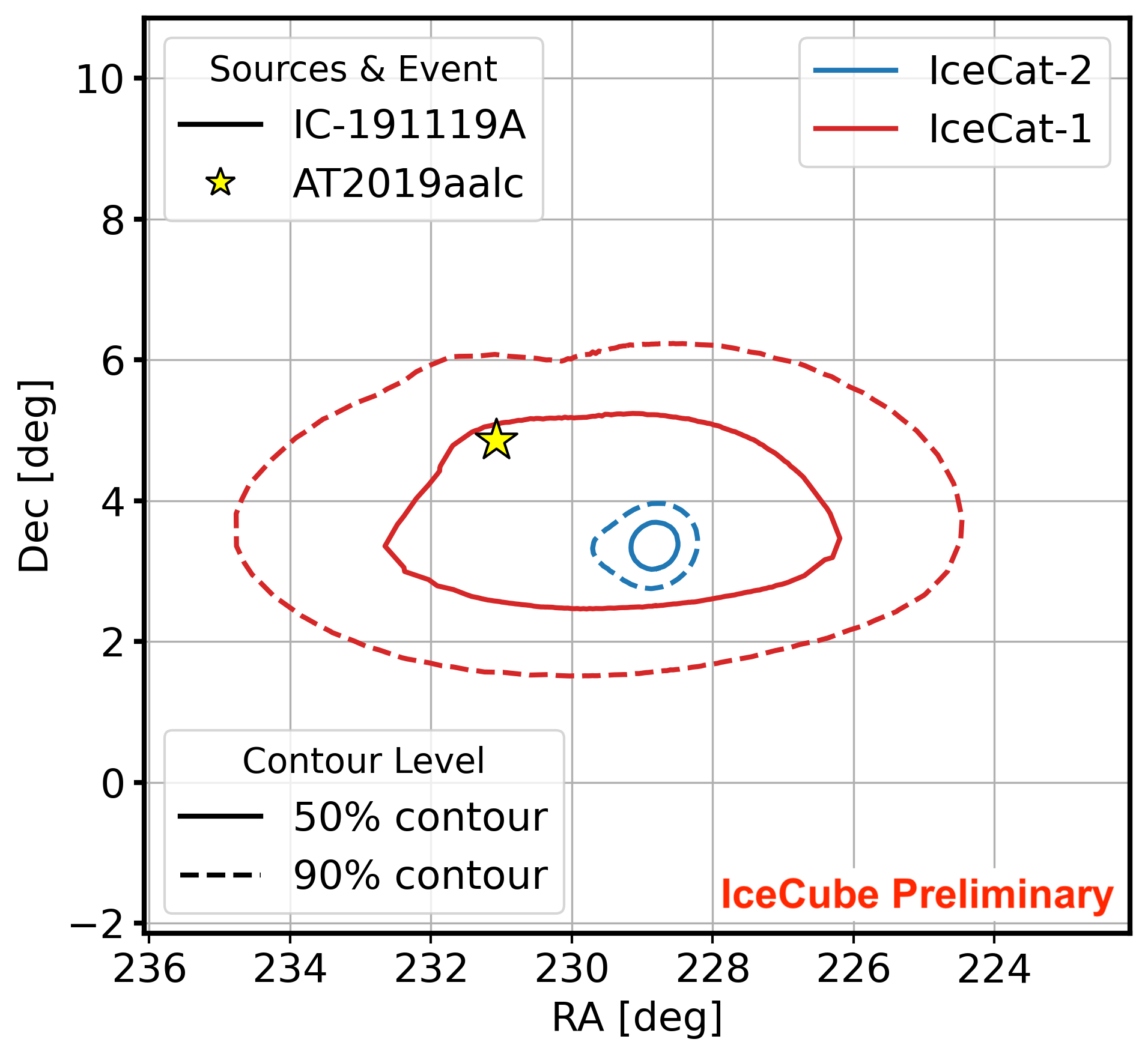}
    \caption{AT2019aalc}
\end{subfigure}

\caption{Sky maps showing spatial coincidences between IceCat track alerts and several TDEs (yellow stars). Each panel highlights improvements in event localization from IceCat-1 (red) to IceCat-2 (blue), with solid and dashed lines showing the 50\% and 90\% containment regions, respectively.}
\label{fig:tde}
\end{figure}

Another notable candidate is the Seyfert galaxy NGC 7469, proposed as a potential neutrino emitter based on its directional coincidence with two IceCube alerts (IC-220424A and IC-230416A) \cite{sommani}. The chance probability of both neutrinos being spatially coincident with NGC 7469 was estimated to correspond to a 3.2$\sigma$ significance. Panels (a) and (b) of Fig.~\ref{fig:ngc} show the position of NGC 7469 relative to IC-220424A and IC-230416A in IceCat-1 and IceCat-2, respectively. While the contour areas around the best-fit directions for the track alerts are significantly reduced in IceCat-2, the source position is still contained within the contours of both neutrino alerts.

Additionally, a few tidal disruption event (TDE) candidates (e.g., AT2019dsg, AT2019fdr, and AT2019aalc) have been associated in recent years with IceCube track alerts through multimessenger follow-up searches \cite{2021NatAs...5..510S,2022PhRvL.128v1101R,van_Velzen_2024}. In each case, the neutrino detection occurred approximately 100 days after the peak of the optical–ultraviolet luminosity. In some of these associations, the TDE positions were located near the edge of the neutrino localization contours, leaving room for uncertainty. Thanks to the improved reconstructions provided by IceCat-2, these associations can now be revisited with greater accuracy. We find that the TDE positions are, in all cases, well outside the updated containment regions, effectively ruling out an association with the corresponding IceCube alerts (see Fig.~\ref{fig:tde}).
\vspace{-4mm}
\section{Search for correlation with potential source candidates for association}
\label{sec:catalog_cross_check}
\vspace{-3mm}
Following the approach previously adopted for IceCat-1, we also re-evaluated the directional correlation of the updated alert tracks with several gamma-ray catalogs (3FHL, 4FGL-DR4, 3HWC, TeVCat) as well as the Swift-BAT X-ray catalog. For each of the 365 alerts in the preliminary IceCat-2 sample, we searched the aforementioned catalogs for sources located within the 90\% uncertainty contour of the alert’s updated reconstructed direction. In addition, to estimate the number of coincidences expected by chance, we randomized the alert directions in right ascension 1000 times and recorded the number of matches for each iteration. For each catalog, we find that the number of coincidences is consistent with the median expectation due to chance correlation (see Tab. \ref{tab}). The reduced number of expected coincidences with respect to IceCat-1 \cite{icecat1} is consistent with the significant improvement in directional reconstruction introduced in IceCat-2.
\begin{table}[t!]
\centering
\small
\begin{tabular}{lcc}
\hline
Catalog  & Observed Coincidences & Expected Coincidences \\
\hline
4FGL-DR4 & 93 & 89 \\
3FHL & 29 & 28 \\
3HWC & 2 & 2 \\
TeVCat & 6 & 5 \\
Swift-BAT & 35 & 32 \\
\hline
\end{tabular}
\caption{Observed number of alerts containing a catalog source within the 90\% error contours, and the number expected from random coincidences.}\label{tab}
\end{table}

\vspace{-4mm}
\section{Summary}
\label{sec:summary}
\vspace{-3mm}
To enhance multi-messenger capabilities, IceCube has developed an improved reconstruction method for its real-time alert system. As a result, the original IceCube Event Catalog of Alert Tracks (IceCat-1) has been revised. IceCat-2 incorporates the new reconstructions, updated calibrations, exclusion of likely CR-induced events, and new alerts since the last release. We present the preliminary IceCat-2 catalog, which contains 365 track-like alerts from May 2011 (IC-110514A) to January 2, 2025 (IC-250102A). Studies of spatial correlations with astrophysical sources will take advantage of IceCat-2 significantly (e.g., ~\cite{new_tdes_searches}). Notably, it improves the angular uncertainty areas by a factor $\sim5$, with median 50\% and 90\% containment areas of 0.4 and 1.3 square degrees, respectively. A public release of the full catalog will follow a peer-reviewed publication, alongside several analyses on source population correlations.

\vspace{-4mm}
\begingroup
\footnotesize
\setlength{\bibsep}{1pt}
\bibliographystyle{ICRC}
\bibliography{references}
\endgroup

\clearpage

\section*{Full Author List: IceCube Collaboration}

\scriptsize
\noindent
R. Abbasi$^{16}$,
M. Ackermann$^{63}$,
J. Adams$^{17}$,
S. K. Agarwalla$^{39,\: {\rm a}}$,
J. A. Aguilar$^{10}$,
M. Ahlers$^{21}$,
J.M. Alameddine$^{22}$,
S. Ali$^{35}$,
N. M. Amin$^{43}$,
K. Andeen$^{41}$
% ...
\\
\\
% ...
$^{10}$ Universit{\'e} Libre de Bruxelles, Science Faculty CP230, B-1050 Brussels, Belgium \\
% ...
$^{16}$ Department of Physics, Loyola University Chicago, Chicago, IL 60660, USA \\
$^{17}$ Dept. of Physics and Astronomy, University of Canterbury, Private Bag 4800, Christchurch, New Zealand \\
% ...
$^{21}$ Niels Bohr Institute, University of Copenhagen, DK-2100 Copenhagen, Denmark \\
$^{22}$ Dept. of Physics, TU Dortmund University, D-44221 Dortmund, Germany \\
% ...
$^{35}$ Dept. of Physics and Astronomy, University of Kansas, Lawrence, KS 66045, USA \\
% ...
$^{39}$ Dept. of Physics and Wisconsin IceCube Particle Astrophysics Center, University of Wisconsin{\textemdash}Madison, Madison, WI 53706, USA \\
% ...
$^{41}$ Department of Physics, Marquette University, Milwaukee, WI 53201, USA \\
% ...
$^{43}$ Bartol Research Institute and Dept. of Physics and Astronomy, University of Delaware, Newark, DE 19716, USA \\
% ...
$^{63}$ Deutsches Elektronen-Synchrotron DESY, Platanenallee 6, D-15738 Zeuthen, Germany \\
$^{\rm a}$ also at Institute of Physics, Sachivalaya Marg, Sainik School Post, Bhubaneswar 751005, India \\
% ...

\subsection*{Acknowledgments}

\noindent
The authors gratefully acknowledge the support from the following agencies and institutions:
USA {\textendash} U.S. National Science Foundation-Office of Polar Programs,
U.S. National Science Foundation-Physics Division,
U.S. National Science Foundation-EPSCoR,
U.S. National Science Foundation-Office of Advanced Cyberinfrastructure,
Wisconsin Alumni Research Foundation,
Center for High Throughput Computing (CHTC) at the University of Wisconsin{\textendash}Madison,
Open Science Grid (OSG),
Partnership to Advance Throughput Computing (PATh),
Advanced Cyberinfrastructure Coordination Ecosystem: Services {\&} Support (ACCESS),
Frontera and Ranch computing project at the Texas Advanced Computing Center,
U.S. Department of Energy-National Energy Research Scientific Computing Center,
Particle astrophysics research computing center at the University of Maryland,
Institute for Cyber-Enabled Research at Michigan State University,
Astroparticle physics computational facility at Marquette University,
NVIDIA Corporation,
and Google Cloud Platform;
Belgium {\textendash} Funds for Scientific Research (FRS-FNRS and FWO),
FWO Odysseus and Big Science programmes,
and Belgian Federal Science Policy Office (Belspo);
Germany {\textendash} Bundesministerium f{\"u}r Forschung, Technologie und Raumfahrt (BMFTR),
Deutsche Forschungsgemeinschaft (DFG),
Helmholtz Alliance for Astroparticle Physics (HAP),
Initiative and Networking Fund of the Helmholtz Association,
Deutsches Elektronen Synchrotron (DESY),
and High Performance Computing cluster of the RWTH Aachen;
Sweden {\textendash} Swedish Research Council,
Swedish Polar Research Secretariat,
Swedish National Infrastructure for Computing (SNIC),
and Knut and Alice Wallenberg Foundation;
European Union {\textendash} EGI Advanced Computing for research;
Australia {\textendash} Australian Research Council;
Canada {\textendash} Natural Sciences and Engineering Research Council of Canada,
Calcul Qu{\'e}bec, Compute Ontario, Canada Foundation for Innovation, WestGrid, and Digital Research Alliance of Canada;
Denmark {\textendash} Villum Fonden, Carlsberg Foundation, and European Commission;
New Zealand {\textendash} Marsden Fund;
Japan {\textendash} Japan Society for Promotion of Science (JSPS)
and Institute for Global Prominent Research (IGPR) of Chiba University;
Korea {\textendash} National Research Foundation of Korea (NRF);
Switzerland {\textendash} Swiss National Science Foundation (SNSF).

\end{document}